\documentclass[english,aps,prx,twocolumn,superscriptaddress,longbibliography]{revtex4-2}

\usepackage{mathrsfs}
\usepackage{upgreek}
 
\usepackage[symbol]{footmisc}
\RequirePackage{amssymb}
\RequirePackage{amsmath}
\RequirePackage{gensymb}
\RequirePackage{graphicx}
\RequirePackage{natbib}
\RequirePackage{nicefrac}
\RequirePackage{multirow}
\RequirePackage{epstopdf}
\RequirePackage{float}
\RequirePackage{physics}
\RequirePackage{braket}
\RequirePackage[linktocpage=true, colorlinks, allcolors=blue]{hyperref}
\RequirePackage{color} 
\RequirePackage{epigraph}
\RequirePackage{breqn}
\RequirePackage{bbold}
\RequirePackage[utf8]{inputenc}
\RequirePackage{newunicodechar}

\newboolean{ShowComments}
\setboolean{ShowComments}{true}  

\ifthenelse{\boolean{ShowComments}}%
{
	\newcommand{\ColorComment}[3]{%
		{\colorbox{#1}{\color{white}   \textsf{\textbf{#2}}} \textcolor{#1}{#3}}}
	\newcommand{\nyacite}[1]{[#1]}
}%
{
	\newcommand{\ColorComment}[3]{}
	\newcommand{\nyacite}[1]{}
}%

\definecolor{jpcolor}{rgb}{0,0,1}
\definecolor{amcolor}{rgb}{0.2,0.5,0.4}

\definecolor{new}{rgb}{.38,.6,.38}
\definecolor{old}{rgb}{1,0,0}

\newcommand{\princeton}{\affiliation{Department of Physics, Princeton University, Princeton, New Jersey~08544, USA}}
\newcommand{\sandia}{\affiliation{Sandia National Laboratories, Albuquerque, New Mexico~87185, USA}}

\begin{document}
\title{Two-qubit silicon quantum processor with operation fidelity exceeding 99\%}

\author{A. R. Mills}
\princeton
\author{C. R. Guinn}
\princeton
\author{M. J. Gullans}
\altaffiliation{Present address: Joint Center for Quantum Information and Computer Science, NIST/University of Maryland, College Park, Maryland~20742, USA}
\princeton
\author{A. J. Sigillito}
\altaffiliation{Present address: Department of Electrical and Systems Engineering, University of Pennsylvania, Philadelphia, Pennsylvania~19104, USA}
\princeton
\author{M. M. Feldman}
\princeton
\author{E. Nielsen}
\sandia
\author{J. R. Petta}
\princeton

\begin{abstract}
Silicon spin qubits satisfy the necessary criteria for quantum information processing. However, a demonstration of high fidelity state preparation and readout combined with high fidelity single- and two-qubit gates, all of which must be present for quantum error correction, has been lacking. We use a two qubit Si/SiGe quantum processor to demonstrate state preparation and readout with fidelity over 97\%, combined with both single- and two-qubit control fidelities exceeding 99\%. The operation of the quantum processor is quantitatively characterized using gate set tomography and randomized benchmarking. Our results highlight the potential of silicon spin qubits to become a dominant technology in the development of intermediate-scale quantum processors.
\end{abstract}

\maketitle
Since the publication of the Loss-DiVincenzo proposal for spin-based quantum information processing in 1998 \cite{Loss1998}, the semiconductor quantum dot research community has worked to satisfy the DiVincenzo criteria for quantum information processing using spin qubits. Electron spins can be initialized and read out using spin-to-charge conversion \cite{Elzerman2004,Petta2005}, single spin qubits can be coherently controlled using oscillating electromagnetic fields \cite{Koppens2006,Nowack2007,Pioro2008}, and nearest neighbor spins can be coherently coupled via the exchange interaction \cite{Petta2005}.

Seminal results for spin qubits were obtained with GaAs quantum dots \cite{Elzerman2004,Petta2005,Koppens2006}, however hyperfine coupling of the electron spin to lattice nuclei greatly limited quantum coherence \cite{Hanson2007_Rev}. A transition to silicon, which can be isotopically enriched, has led to a several order-of-magnitude increase in electron spin coherence times \cite{Tyryshkin2012} as well as improved quantum control fidelities \cite{Yoneda2018,Xue2019,Huang2019}. The small $\sim$100 nm scale of quantum dot spin qubits and the significant capabilities of the silicon microelectronics industry could allow for scaling to system sizes that are capable of fault tolerant operation.

While high single- and two-qubit gate fidelities have been demonstrated in silicon \cite{Yoneda2018,Huang2019,Xue2021,Noiri2021}, state preparation and measurement (SPAM) errors have generally hovered around $\sim$10-20\%. Here we demonstrate a spin-based two qubit quantum processor with all-around high performance fidelities (readout $F>97\%$, simultaneous single qubit control $F>99\%$, and a two-qubit controlled-phase (CPHASE) gate $F>99.8\%$). Our two-qubit gate fidelity exceeds recent reports on spin qubits \cite{Xue2021,Noiri2021} and is competitive with superconducting qubits \cite{Mundada2019,Sung2021}.

High-fidelity quantum control and readout are achieved in the first two qubits (Q1 and Q2) of a six qubit linear array [Fig. 1A]. Quantum dot electrons are vertically confined in an isotopically enriched (800 ppm) $^{28}$Si quantum well and lateral confinement is achieved using an overlapping aluminum gate stack \cite{Zajac2016}. Figure 1B depicts the double quantum dot formed under gates P1 and P2 with the exchange interaction controlled by gate B2 \cite{Petta2005}.  An external magnetic field $B_{E}$ = 365 mT is applied in the plane of the quantum well to Zeeman split the spin-states. The external field adds to the magnetic field generated by a Co micromagnet resulting in electron spin resonance frequencies $f_1$ = 18.247 GHz and $f_2$~=~17.851~GHz.

We first demonstrate high visibility readout and single qubit control fidelities. Single qubit gates are achieved at the symmetric operating point in the (1,1) charge state using electric dipole spin resonance in the transverse field gradient created by the micromagnet \cite{Pioro2008,Reed2016,Martins2016}.  Here $(N_1,N_2)$ denotes the charge occupation of dots 1 and 2. Qubit state preparation and readout are achieved by spin dependent tunneling with the reservoir \cite{Elzerman2004,Veldhorst2015,SOM}. Figure 1C shows Rabi chevrons for each qubit, obtained by measuring the spin-up probability $P_{\uparrow}$ as a function of drive frequency $f$ and microwave burst length $\tau_R$. Rabi oscillations approaching unit visibility are achieved when driving each qubit on resonance [Figure 1D].

\begin{figure*}[t]
	\centering
	\includegraphics[width=2\columnwidth]{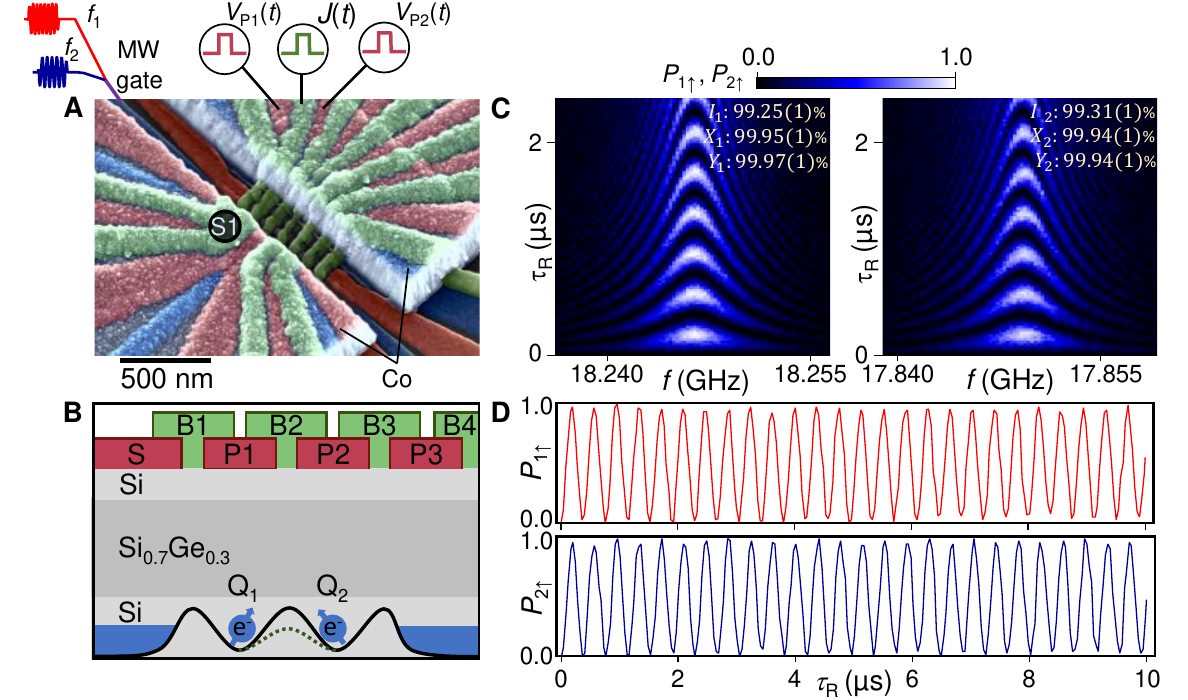} 
	\caption{\textbf{High-fidelity operation of a two-qubit quantum processor.} (\textbf{A}) False-color scanning electron microscope image of the device. Spins are selectively driven in a field gradient of a Co micromagnet using the microwave (MW) gate.	(\textbf{B}) Two electron spins (Q$_1$ and Q$_2$) are trapped beneath gates P1 and P2, and the exchange coupling between the spins is set by the barrier gate B2. (\textbf{C}) Spin-up probabilities, $P_{1\uparrow}$($P_{2\uparrow}$), for Q1(Q2) plotted as a function of drive frequency $f$ and microwave burst length $\tau_R$. Insets: Single qubit gate fidelities extracted from GST. (\textbf{D}) Spin-up probabilities for each qubit when driven on resonance.}
	\label{fig:1}
\end{figure*}

We perform the quantum characterization, verification, and validation protocol of gate set tomography (GST) on the single qubit gates - identity (idle qubit $i$ for $\pi/2$ gate time 70 ns) $I_i$, $\pi_X/2$ rotation $X_i$, and  $\pi_Y/2$ rotation $Y_i$ - and estimate single qubit control fidelities above 99.9\% when driving and measuring one qubit at a time [Fig. 1C, inset] \cite{Nielsen2021}. The fidelities are limited by decoherence [$T_2^*(T_2)$ = 1.7(23) $\mu$s and 2.3(102) $\mu$s for Q1 and Q2] and are comparable to the highest single spin qubit gate fidelities in the literature \cite{Yoneda2018,Yang2019}. SPAM errors extracted from GST are quantified by the initialization fidelities $\rho_{0,1}=99.4\%$ and $\rho_{0,2}=97.5\%$ and the measurement fidelities $M_1 = 98.1\%$ and $M_2 = 99.8\%$, making the overall operation fidelity high enough to support common error correction protocols.

\begin{figure*}[t]
	\centering
	\includegraphics[width=2\columnwidth]{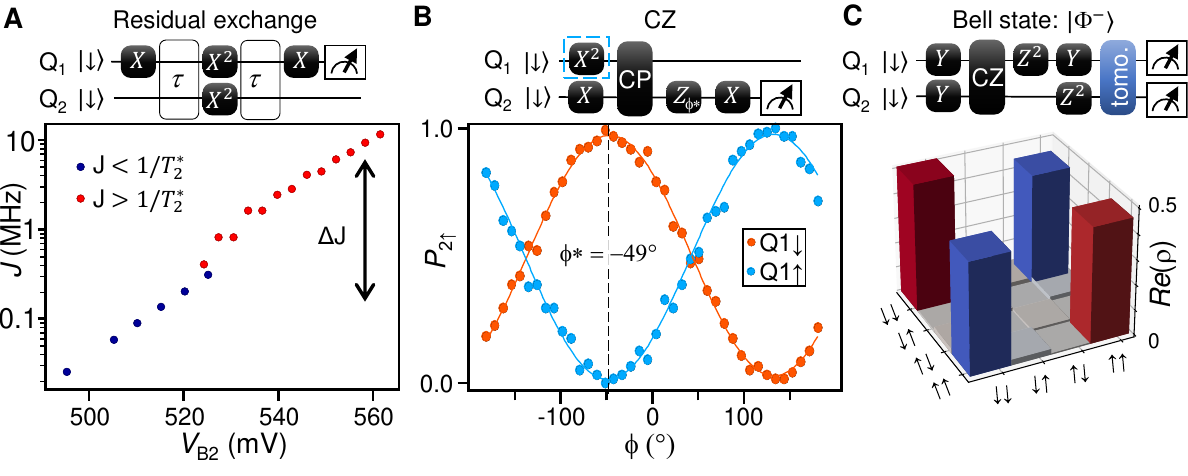} 
	\caption{\textbf{Optimization of the CZ gate.} (\textbf{A}) Exchange interaction measured as a function of $V_{B2}$ using a combination of Ramsey and Hahn echo type sequences (see text). The quantity $\Delta J$ indicates how far we can dynamically tune exchange. (\textbf{B}) The CZ is tuned by preparing the target spin in a superposition state, applying the C-Phase operation at $J$$\sim$5 MHz for 40 ns, and then adjusting the phase $Z_{\phi}$ such that $P_{2\uparrow}$ = 1 ($P_{2\uparrow}$ = 0) when Q1 is prepared in the spin-down (spin-up) state using an optional $\pi$-pulse (blue dashed box). (C) Bell state tomography of the $\ket{\Phi^-}$ state yields the extracted density matrix (shown) with a raw state fidelity $F_u$ = 96.3\%. Correcting for SPAM errors yields a fidelity $F_c$ = 97.4\%.}
	\label{fig:2}
\end{figure*}

Building to the two qubit space, we use GST to characterize the qubit control fidelities when both qubits are operated simultaneously ($X_{1} \otimes X_{2}, Y_{1} \otimes X_{2},$ etc.) by combining microwave control signals on the drive gate [MW gate in Fig. 1A]. GST estimates an average simultaneous single qubit control fidelity $F$ = 99.4 $\pm$ 0.1\% when operating in the two qubit space. Finally, we cross-check these GST results using the widely accepted protocol of randomized benchmarking (RB) and achieve $F_1$ = 99.50 $\pm$ 0.02\%, $F_2$ = 99.48 $\pm$ 0.02\%, and a joint fidelity of 99.13 $\pm$ 0.03\%. These results build significantly on past attempts to simultaneously drive spin qubits, where single qubit control fidelities were as low as 97\% \cite{Xue2019}.

\begin{figure}[t]
	\centering
	\includegraphics[width=\columnwidth]{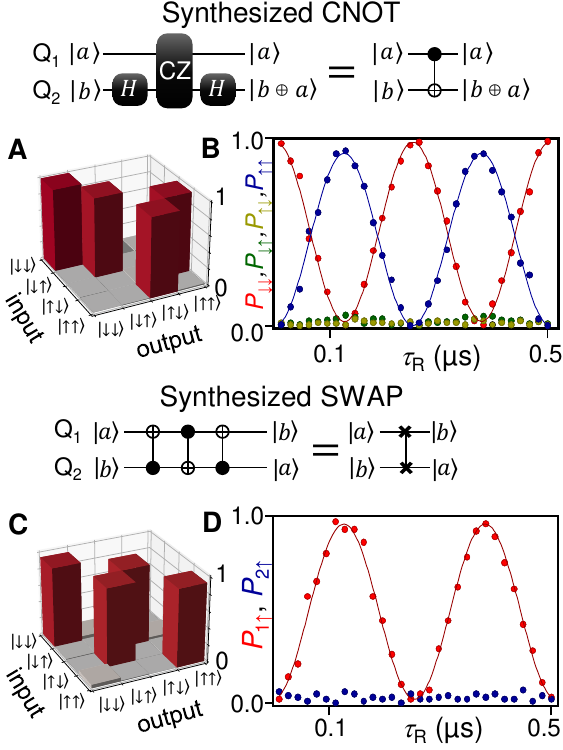} 
	\caption{\textbf{Synthesis of CNOT and SWAP gates using the primitive CZ gate.} (\textbf{A}) 	Input-output table illustrating the effect of the synthesized CNOT gate on the four different input product states ($\ket{\uparrow\uparrow}, \ket{\uparrow\downarrow},\ket{\downarrow\uparrow},\ket{\downarrow\downarrow}$). (\textbf{B}) Joint state probabilities  $P_{\uparrow\uparrow}$, $P_{\uparrow\downarrow}$, $P_{\downarrow\uparrow}$, and $P_{\downarrow\downarrow}$ plotted as a function of microwave burst length $\tau_R$ at frequency $f_2$ showing that the joint state $\ket{\downarrow\downarrow}$ is anticorrelated with $\ket{\uparrow\uparrow}$. (\textbf{C}) Input-output table for the synthesized SWAP gate. (\textbf{D}) $P_{1\uparrow}$ and $P_{2\uparrow}$ measured as a function of Q2 drive time $\tau_R$ at frequency $f_2$ following a SWAP of the Q2 state onto Q1.
	}
	\label{fig:3}
\end{figure}

Two qubit control is achieved by pulsing on gate B2 to turn on the exchange interaction $J(V_{B_2})$ \cite{Petta2005}, which up to single qubit rotations yields a controlled-Z (CZ) gate $U_{CZ}$ = diag(1,1,1,-1) in the regime where the magnetic field gradient exceeds exchange $\Delta E_z \gg J$ \cite{Meunier2011,Russ2018}. Figure 2A demonstrates a 3-decade variation of $J(V_{B2})$. Exchange is extracted in the high-$J$ regime ($J>1/T_2^*$) by measuring time domain exchange oscillations \cite{Watson2018}, whereas a spin echo is utilized to measure residual exchange down to a $T_2$ limit of ~10 kHz [blue in Fig. 2A]. In Fig. 2B we optimize the CZ gate by preparing the target qubit (Q2 here) in a superposition state and the control qubit in either $\uparrow$ or $\downarrow$. Exchange is then pulsed on for 40 ns using a smoothed square pulse \cite{SOM} to execute a C-Phase gate. Afterwards, a software Z rotation $Z_\phi$ is applied to realize a CZ gate. As an initial demonstration of full two-qubit control with low SPAM errors, we perform Bell state tomography and use maximum likelihood estimation to achieve the Bell state fidelities $\ket{\Psi^+}=97.5\%(98.3\%)$, $\ket{\Psi^-}=97.0\%(98.3\%)$, $\ket{\Phi^+}=95.4\%(97.2\%)$, $\ket{\Phi^-}=96.3\%(97.4\%)$, without(with) SPAM corrections included [Fig. 2C] \cite{SOM}. These results significantly improve upon past measurements with SPAM corrected fidelities of 78-90\% \cite{Zajac2018,Watson2018,Huang2019,Leon2021} and are comparable to the \textit{simulated} Bell state fidelities obtained by Xue \textit{et al.} \cite{Xue2021}.

To demonstrate integrated control of the two qubit processor, we combine the CZ with other primitive qubit operations to create familiar two qubit gates (e.g. CNOT and SWAP). We first synthesize a CNOT gate using the Hadamard and CZ gates. Figure 3A shows the raw input-output measurement results of performing the synthesized CNOT gate on the four different input product states with Q2 as the target. To show the target qubit will follow the control qubit state we prepare Q1 in the $\downarrow$ state and Q2 in an arbitrary superposition state using a Rabi drive pulse for time $\tau_R$. This state preparation routine is followed by the CNOT with Q1 as the target qubit this time. The result is high visibility anti-correlated oscillations of the $\downarrow\downarrow$ and $\uparrow\uparrow$ joint state probabilities, Fig. 3B. Additionally, we generate synthesized SWAP gates with three alternating CNOT operations and measure the input-output SWAP table, Fig. 3C. To show a SWAP of an arbitrary superposition state $\ket{\psi}$ we perform a Rabi drive pulse on Q2 and then SWAP the state onto Q1, resulting in Q2 being in the $\downarrow$ state and Q1 in the $\ket{\psi}$ state [Fig. 3D].

We turn to two qubit interleaved RB to quantify the overall performance of our processor \cite{Magesan2011}. The two qubit Clifford group $C_2$ in this experiment has 576 single-qubit elements and integrates the CZ into 10944 two-qubit elements containing CNOT-, iSWAP-, and SWAP-like operations similar to those demonstrated in Fig.~3. With this technique, the CZ and synthesized CNOT fidelities can be determined by interleaving these operations in the benchmarking sequences and comparing to the reference curve. To thoroughly sample the Clifford group and obtain an accurate estimate of our CZ and CNOT fidelities we randomize 125 unique sequences for each reference and interleaved measurement going to sequence lengths as long as 65 total Clifford operations and average 160 times [Fig.~4]. The resulting two qubit RB fidelities for the CZ and CNOT are $F_{CZ}=99.81\pm0.17\%$ and $F_{CNOT}=98.62\pm0.16\%$ with error bars determined by bootstrapping \cite{Barends2014}. 

We have demonstrated full two qubit control in a silicon quantum device with simultaneous single qubit control fidelities exceeding 99\% and a primitive two-qubit CZ-gate fidelity exceeding 99.8\%. In contrast to previous implementations \cite{Zajac2018,Watson2018,Noiri2021,Xue2021}, SPAM errors are very low $<$3\%. Our demonstration represents the highest total operation fidelity in a two qubit processor realized in silicon quantum dots with performance capable of fault tolerant operation \cite{Fowler2012}. These experiments demonstrate two qubit gates with silicon spin qubits at speeds exceeding trapped ions \cite{Bruzewicz2019} and fidelities comparable with superconducting qubits \cite{Mundada2019,Sung2021}. Given recent advances in quantum dot fabrication \cite{Zajac2016,Ha2021} spin qubits are poised to scale-up to larger multi-qubit quantum processors.

\begin{figure}[tbp]
	\centering
	\includegraphics[width=\columnwidth]{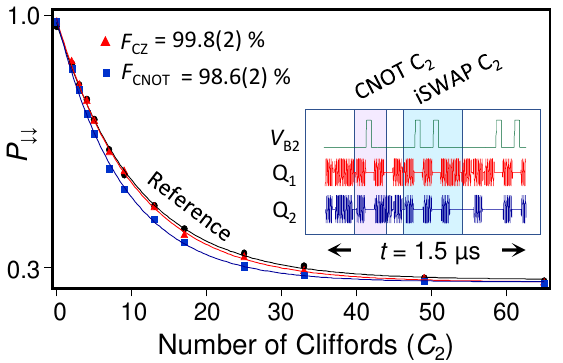}
	\caption{\textbf{Two qubit interleaved RB.} Return probability $P_{\downarrow\downarrow}$ as a function of the number of two-qubit Clifford operations. Reference data are shown in black with error rate $r_{ref}=0.0679$ \cite{Barends2014}. Interleaved RB yields a two-qubit CZ fidelity $F_{\rm CZ}$ = 99.8\% and synthesized CNOT fidelity $F_{\rm CNOT}$ = 98.6\%. Inset: Small portion of an RB sequence that includes a CNOT-like Clifford followed by an iSWAP-like Clifford. }
	\label{fig:RB}
\end{figure}

\begin{acknowledgements}
We thank Remy Delva for computational support and David Zajac for assistance with sample fabrication. The Si/SiGe heterostructure used in these experiments was provided by HRL Laboratories, LLC. Supported by Army Research Office grant W911NF-15-1-0149 and DARPA grant D18AC0025. Devices were fabricated in the Princeton University Quantum Device Nanofabrication Laboratory, which is managed by the department of physics. The authors acknowledge the use of Princeton's Imaging and Analysis Center, which is partially supported by the Princeton Center for Complex Materials, a National Science Foundation MRSEC program (DMR-2011750).
The Sandia National Laboratories part of this work was funded in part the by the U.S. Department of Energy, Office of Science, Office of Advanced Scientific Computing Research’s Quantum Testbed Pathfinder. Sandia National Laboratories is a multimission laboratory managed and operated by National Technology and Engineering Solutions of Sandia, LLC, a wholly owned subsidiary of Honeywell International, Inc., for the U.S. Department of Energy’s National Nuclear Security Administration under contract DE-NA0003525.  All statements of fact, opinion or conclusions contained herein are those of the authors and should not be construed as representing the official views or policies of IARPA, the ODNI, the U.S. Department of Energy, or the U.S. Government.
\end{acknowledgements}

\bibliography{PettaLab_refs2}

\end{document}